\documentclass[preprint,5p]{elsarticle}
\usepackage{graphicx,epsfig,amssymb,times}
\usepackage{amsmath,amsfonts,amssymb}
\usepackage{bm}
\usepackage{bbold}
\usepackage{epstopdf}
\usepackage[linktocpage,colorlinks]{hyperref}
\usepackage[caption=false]{subfig}
\usepackage[usenames]{color}  
\journal{Physics Letter B}
\usepackage{graphicx,epsfig,amssymb} %include figure files
\usepackage{amsmath,amsfonts, times}
\usepackage{bm} %include bold math: \bm{} creates bold letters in math mode

\usepackage{epstopdf}
\usepackage[linktocpage,colorlinks]{hyperref}
\usepackage[caption=false]{subfig}
\usepackage[usenames]{color}     
\usepackage{natbib}
\usepackage{soul}
\usepackage[utf8x]{inputenc}

\definecolor{coolblack}{rgb}{0.0, 0.18, 0.39}
\definecolor{darkred}{rgb}{0.5,0,0}
\definecolor{darkgreen}{rgb}{0,0.5,0}
\definecolor{darkblue}{rgb}{0,0,0.5}
\definecolor{lapislazuli}{rgb}{0.15, 0.38, 0.61}
\definecolor{venetianred}{rgb}{0.78, 0.03, 0.08}
\definecolor{bleudefrance}{rgb}{0.19, 0.55, 0.91}
\definecolor{dogwoodrose}{rgb}{0.84, 0.09, 0.41}
\hypersetup{colorlinks=true, citecolor=darkgreen, linkcolor=darkblue, 
	urlcolor = blue}

\def\be{\begin{equation}}
\def\ee{\end{equation}}

\newcommand{\bea}{\begin{eqnarray}}
\newcommand{\eea}{\end{eqnarray}}
\newcommand{\ben}{\begin{enumerate}}
	\newcommand{\een}{\end{enumerate}}
\newcommand{\bi}{\begin{itemize}}
	\newcommand{\ei}{\end{itemize}}
\newcommand{\Tr}{\mbox{Tr}}

\newcommand{\bqa}{\begin{eqnarray}}
\newcommand{\eqa}{\end{eqnarray}}

\def\ga{\mathrel{\raise.3ex\hbox{$>$\kern-.75em\lower1ex\hbox{$\sim$}}}}
\def\la{\mathrel{\raise.3ex\hbox{$<$\kern-.75em\lower1ex\hbox{$\sim$}}}}

\def\be{\begin{equation}}
\def\ee{\end{equation}}

\def\Tr{{\rm Tr}}
\def\I_M{{I_{\scriptscriptstyle M\times M}}}

\def\be{\begin{equation}}
\def\ee{\end{equation}}
\def\bea{\begin{eqnarray}}
\def\eea{\end{eqnarray}}
\newcommand{\beq}{\begin{eqnarray}}
\newcommand{\eeq}{\end{eqnarray}}

\usepackage{graphicx,epsfig,amssymb} %include figure files
\usepackage{amsmath,amsfonts, times}
\usepackage{bm} %include bold math: \bm{} creates bold letters in math mode

\usepackage{epstopdf}
\usepackage[linktocpage,colorlinks]{hyperref}
\usepackage[caption=false]{subfig}
\usepackage[usenames]{color}     
\usepackage{natbib}
\usepackage{soul}
\usepackage[utf8x]{inputenc}

\definecolor{coolblack}{rgb}{0.0, 0.18, 0.39}
\definecolor{darkred}{rgb}{0.5,0,0}
\definecolor{darkgreen}{rgb}{0,0.5,0}
\definecolor{darkblue}{rgb}{0,0,0.5}
\definecolor{lapislazuli}{rgb}{0.15, 0.38, 0.61}
\definecolor{venetianred}{rgb}{0.78, 0.03, 0.08}
\definecolor{bleudefrance}{rgb}{0.19, 0.55, 0.91}
\definecolor{dogwoodrose}{rgb}{0.84, 0.09, 0.41}
\hypersetup{colorlinks=true, citecolor=darkgreen, linkcolor=darkblue, 
	urlcolor = blue}

\def\be{\begin{equation}}
\def\ee{\end{equation}}

\def\be{\begin{equation}}
\def\ee{\end{equation}}

\def\be{\begin{equation}}
\def\ee{\end{equation}}
\def\bea{\begin{eqnarray}}
\def\eea{\end{eqnarray}}

\begin{document}
\begin{frontmatter}
\title{QCD at finite isospin density: chiral perturbation theory confronts
lattice data}
\author[a,b]{Prabal Adhikari}
%\ead{pa100@wellesley.edu}
\author[b,c]{Jens O. Andersen\corref{mycorrespondingauthor}}
\address[a]{Wellesley College, Department of Physics, 106 Central Street,
  Wellesley, MA 02481, United States}
\address[b]{Department of Physics, 
Norwegian University of Science and Technology, H{\o}gskoleringen 5,
N-7491 Trondheim, Norway}
\address[c]{Niels Bohr International Academy, Blegdamsvej 17,
DK-2100 Copenhan, Denmark}
\cortext[mycorrespondingauthor]{Corresponding author}
\ead{andersen@tf.phys.ntnu.no}
\begin{abstract}
  We consider the thermodynamics of three-flavor QCD in the pion-condensed
  phase at nonzero isospin chemical potential ($\mu_I$) and vanishing
  temperature
  using chiral perturbation theory in the isospin limit.
  The transition from the vacuum phase to a superfluid phase with a
  Bose-Einstein condensate of charged pions is shown to be second order
  and takes place at $\mu_I=m_{\pi}$.
  We calculate the pressure, isospin density, and energy density
  to next-to-leading order in the low-energy expansion.
  Our results are compared with recent high-precision lattice simulations
  as well as previously obtained results in two-flavor chiral perturbation
  theory. The agreement between the lattice results and the predictions from
  three-flavor chiral perturbation theory is very good for
  $\mu_I<200$ MeV. For larger values of $\mu_I$, the agreement between
  lattice data and the two-flavor predictions is surprisingly good
  and better than with the three-flavor predictions.
    Finally, in the limit $m_{s}\gg m_{u}=m_{d}$, we show
    that the three-flavor observables  reduce to the
    two-flavor observables
    with renormalized parameters. The disagreement between the
    results for two-flavor and three-flavor $\chi$PT can
    largely be explained
  by the differences in the measured low-energy constants.
\end{abstract} 


\end{frontmatter}

%%%%%%%%%%%%%%%%%%%%%%%%%%%%%%%%%%%%%%%%%%%%%%
\section{Introduction}\label{sec:int}
%%%%%%%%%%%%%%%%%%%%%%%%%%%%%%%%%%%%%%%%%%%%%%
QCD in extreme conditions, i.e. high temperature and density has received
a lot of attention in the past decades due to its relevance to the 
early universe, heavy-ion collisions, and compact
stars~\cite{raja,alford,fukurev}.
For example, QCD at finite baryon density ($\mu_B$)
is of significant interest since the equation
of state (EoS) is used as input for calculating the macroscopic properties
of neutron stars. However,
lattice QCD cannot be applied to QCD at nonzero baryon
density due to the sign problem: integrating out the fermions in the
path integral for the partition function gives rise to a functional determinant
that can be considered part of the probability measure. At $\mu_B\neq0$,
this determinant is complex and standard Monte Carlo techniques cannot
be applied. A way to circumvent this problem, for high temperatures and
small chemical potentials, is by Taylor expanding the thermodynamic
quantities about zero $\mu_B$~\cite{borsa}. For small $T$ and large
$\mu_B$, this is obviously hopeless.
Due to asymptotic freedom, we expect to be able to
use weak-coupling techniques at very high densities~\cite{freed1,freed2}.
In the weak-coupling expansion the series is now known to order
$\alpha_s^2$ for massive quarks~\cite{finland} and
$\alpha_s^3\log^2\alpha_s$ for massless quarks~\cite{finland2}.
For lower densities, where weak-coupling techniques do not apply, we have
to use low-energy models of QCD, see Ref.~\cite{gordon}
for a recent review.

There are variants of QCD
that do not suffer from the sign problem. These
include two-color QCD~\cite{cotter},
three-color QCD with fermions in the adjoint
representation~\cite{simon}, zero density
QCD in an external magnetic field~\cite{bruck},
and three-color QCD at finite
isospin~\cite{kogut1,kogut2,gergy1,gergy2,gergy3}. The absence of
the sign problem implies that one can simulate these systems on the lattice
and compare the results with low-energy models and theories.
In the case of QCD at finite isospin chemical potential, one finds at $T=0$, a
transition
from the vacuum
to a pion-condensed phase at a critical isospin chemical potential
$\mu_I^c=m_{\pi}$.
The mechanism of pion condensation and the transition to a pion superfluid
phase out of the vacuum is simply that it is energetically favorable
to form such a condensate for $\mu_I\geq\mu_I^c$.
Moreover, with increasing isospin chemical potential, it is
expected that there is a crossover to a BCS phase.
Since the order parameter in the BCS phase has the same quantum numbers
as a charged pion condensate, this is not a true phase transition, but
associated with the formation of a Fermi surface and subsequent
condensation of Cooper pairs. A very recent review on meson condensation
can be found in Ref.~\cite{mannarev}.

Chiral perturbation theory ($\chi$PT) is a low-energy effective theory of QCD
based only on its global
symmetries and the degrees of freedom, and 
the predictions of $\chi$PT are, therefore, model
independent~\cite{wein,gasser1,gasser2,bein}.
It has been remarkably successful in describing the phenomenology of
the pseudo-Goldstone bosons that result from the spontaneous breakdown
of chiral symmetry  %$SU(3)_L\times SU(3)_R$ symmetry to $SU(3)_V$
in the QCD vacuum. 
$\chi$PT at finite isospin was first considered by Son and Stephanov
in their seminal paper two decades ago~\cite{son}, in which all the
leading order results were derived.

In this letter, we calculate the effective potential in chiral perturbation
theory 
at next-to-leading (NLO) order in the low-energy expansion for
three flavors at finite isospin chemical potential.
While the phase diagram as functions of isospin and strange chemical potentials
($\mu_S$)
has been mapped out and leading order (LO) thermodynamic functions have been
known for two decades~\cite{son,kogutchpt},
the leading quantum corrections at finite $\mu_I$
are presented here for the first time, however, see
Ref.~\cite{split2} for some partial NLO results in two-color QCD and
Refs.~\cite{loewe,fragaiso,cohen2,janssen,carig0,carigchpt,luca}
for various aspects of $\chi$PT for three-color QCD
including some NLO effects.
Finite isospin systems have also been studied in the
context of low-energy effective models including the non-renormalizable
Nambu-Jona-Lasinio model~\cite{2fbuballa,toublannjl,bar2f,he2f,heman2,heman,ebert1,ebert2,sun,lars,2fabuki,heman3,he3f,ricardo,ruggi}, 
and the
renormalizable quark-meson model~\cite{lorenz,ueda,qmstiele,allofus}.

We derive the pressure, isospin density, and equation of state,
and compare these quantities with recent lattice results as well
earlier results from two-flavor $\chi$PT~\cite{usagain}.
  In the large-$m_s$ limit, the three-flavor result is matched onto the
two-flavor result of Ref.~\cite{usagain} with renormalized parameters.
The disagreement between the two-flavor and three-flavor results
are discussed and shown to be related to the differences in the
experimental values of the low-energy constants.
Results on the thermodynamics of the kaon-condensed phases at
finite $\mu_S$ and $\mu_I$
as well as calculational details can be found in an accompanying long
paper~\cite{moreresults}.

\section{Chiral perturbation theory}
As mentioned above, $\chi$PT is an effective low-energy theory of QCD
based solely on its global
symmetries and low-energy degrees of freedom. In massless
three-flavor QCD,
the symmetry is $SU(3)_L\times SU(3)_R\times U(1)_B$, which in the
vacuum is broken down to $SU(3)_V\times U(1)_B$.
For two degenerate light quarks, the symmetry
is $SU(2)_I\times U(1)_Y\times U(1)_B$. If we add a quark chemical potential
for each flavor, the symmetry is $U(1)_{I_3}\times U(1)_Y\times U(1)_B$.
In three-flavor QCD, we keep the octet of mesons, which implies
that chiral perturbation theory is not valid for arbitrarily large
chemical potential. Considering the hadron spectrum,
one naively expects that the expansion is valid for
$|\mu_u|=|\mu_d|<300$ MeV ~\cite{kogutchpt}. 	
$\chi$PT has a well defined power counting scheme, where
each derivative as well as each factor of a quark mass counts as one power
of momentum $p$. At leading order in momentum, ${\cal O}(p^2)$, there
are only two terms in the chiral Lagrangian
\bqa
\mathcal{L}_{2}=\frac{f^{2}}{4}{\rm Tr}
\left [\nabla_{\mu} \Sigma^{\dagger} \nabla^{\mu}\Sigma 
    \right ]
+{f^2\over4}{\rm Tr}
\left [ \chi^{\dagger}\Sigma+\chi\Sigma^{\dagger}\right ]\; ,
\label{lag0}
\eqa
where $f$ is the bare pion decay constant, $\chi=2B_0M$, 
\bqa
M={\rm diag}(m_u,m_d,m_s)
\eqa
is the quark mass matrix and
$\Sigma=U\Sigma_0U$, where $U=\exp{i\lambda_i\phi_i\over2f}$
and $\Sigma_0=\mathbb{1}$ is the vacuum.
Here $\lambda_i$ are the Gell-mann matrices that satisfy
$\Tr\lambda_i\lambda_j=2\delta_{ij}$ and $\phi_i$ are the
fields parametrizing the Goldstone manifold ($i=1,2...,8$).
In the remainder we work in the isospin limit, $m=m_u=m_d$.
The covariant derivative and its Hermitian conjugate at nonzero quark chemical potentials, $\mu_q$ ($q=u,d,s$), are defined as follows
\bqa
\nabla_{\mu} \Sigma&\equiv&
\partial_{\mu}\Sigma-i\left [v_{\mu},\Sigma \right]\;,\\ 
\nabla_{\mu} \Sigma^{\dagger}&=&
\partial_{\mu}\Sigma^{\dagger}-i [v_{\mu},\Sigma^{\dagger} ]
\;,
\eqa
with
\begin{align}\nonumber
v_{\mu}&=\delta_{\mu0}\ {\rm diag}(\mu_u,\mu_d,\mu_s)
\\ 
&=\delta_{\mu0}\ {\rm diag}(\mbox{$1\over3$}\mu_B
+\mbox{$1\over2$}\mu_I,\mbox{$1\over3$}\mu_B-\mbox{$1\over2$}\mu_I,
\mbox{$1\over3$}\mu_B-\mu_S)\;,
\end{align}
where
$\mu_B={3\over2}(\mu_u+\mu_d)$, 
$\mu_I=\mu_u-\mu_d$, and 
$\mu_S={1\over2}(\mu_u+\mu_d-2\mu_s)$.
It turns out that the Lagrangian is independent of $\mu_B$ which
reflects the fact that all degrees of freedom, namely the meson octet,
have zero baryon number.
Since we are focusing on pion condensation and want to
compare with lattice data, 
we set $\mu_S=0$ such that $v_0={1\over2}\mu_I\lambda_3$.
By expanding the Lagrangian (\ref{lag0}) to second order in the fields,
we obtain the terms needed for our NLO
calculation.~\footnote{A covariant derivative and a mass term 
both count as order $p$ in the low-energy expansion.}

Based on the two-flavor case~\cite{son}, 
the ground state in the pion-condensed phase
is parametrized as~\cite{moreresults}
\begin{align}
\Sigma_{\alpha}=e^{i\alpha(\hat{\phi}_1\lambda_1+\hat{\phi}_2\lambda_2)}=
\cos\alpha+(i\hat{\phi}_1\lambda_1+\hat{\phi}_2\lambda_2)\sin\alpha\;,
\end{align}
where $\alpha$ is a rotation angle
and $\hat{\phi}_1^2+\hat{\phi}_2^2=1$
to ensure that the ground state is normalized,
$\Sigma_{\alpha}^{\dagger}\Sigma_{\alpha}=\mathbb{1}$.
From Eq.~(\ref{lag0}), we find the static Hamiltonian
\bqa\nonumber
{\cal H}_2^{\rm static}&=&
{f^2\over4}\Tr[v_0,\Sigma_{\alpha}][v_0,\Sigma_{\alpha}^{\dagger}]
\\ &&
-{f^2\over2}B_0\Tr[M\Sigma_{\alpha}+M\Sigma_{\alpha}^{\dagger}]\;,
\label{comp}
\eqa
where the first term can be written as
${1\over4}f^2\Tr[v_0,\Sigma_{\alpha}][v_0,\Sigma_{\alpha}^{\dagger}]
={1\over8}f^2\mu_I^2
\Tr[\lambda_3\Sigma_{\alpha} \lambda_3\Sigma_{\alpha}^{\dagger}-\lambda_3^2]$.
There is a competition between the two
terms in Eq.~(\ref{comp}): The first term favors
$\Sigma_{\alpha}$ in the
$\lambda_1$ and $\lambda_2$ directions, while $\Sigma_{\alpha}$
in the second terms prefers the normal vacuum,
$\mathbb{1}$~\cite{son}.
It turns out the that the former only depends on
$\hat{\phi}_1^2+\hat{\phi}_2^2$
and so we choose $\hat{\phi}_2=1$ without loss of generality.
The matrix $\lambda_2$ generates the rotations and the
rotated vacuum is given
by $\Sigma_{\alpha}=A_{\alpha}\Sigma_0A_{\alpha}$
where $A_{\alpha}=e^{i{\alpha\over2}\lambda_2}$, and $\Sigma_0=\mathbb{1}$.
The rotated vacuum can then be written in the form
\bqa\nonumber
  \Sigma_{\alpha}&=&{1+2\cos\alpha\over3}
  +i\lambda_2\sin\alpha+{\cos\alpha-1\over\sqrt{3}}\lambda_8\;\\
%  &=&A_{\alpha}^{\pi}\Sigma_0A_{\alpha}^{\pi}
&=&\begin{pmatrix}
\cos\alpha&\sin\alpha&0\\
-\sin\alpha&\cos\alpha&0\\
0&0&1
\end{pmatrix}
  \;.
  \eqa
Here the rotation in the subspace of the $u$ and the $d$-quark is evident  
and at tree level, we have
$\langle\bar{\psi}\psi\rangle^2+\langle\pi^+\rangle^2=
\langle\bar{\psi}\psi\rangle_{\rm vac}^2$, i.e.
the quark condensate is rotated into
a pion condensate.
%This interpretation does not hold beyond leading order~\cite{usagain}.

The fluctuations around the
condensed or rotated vacuum must also be parametrized and this requires
some care. Naively, one would write the field as $\Sigma=U\Sigma_{\alpha}U$,
where $U=\exp{i\lambda_i\phi_i\over2f}$. However, this
parametrization is incorrect since it can be shown that one cannot
renormalize the effective potential at next-to-leading order using
the standard renormalization of the low-energy couplings appearing
in the NLO Lagrangian. One way of understanding the failure
of this parametrization is to realize that the generators of the
fluctuations about the ground state must also be rotated since the vacuum
itself has been rotated. The field must therefore be written as
\bqa
\Sigma&=&L_{\alpha}\Sigma_{\alpha}R_{\alpha}^{\dagger}\;,
\eqa
where $L_{\alpha}=A_{\alpha}UA_{\alpha}^{\dagger}$ and
$R_{\alpha}=A_{\alpha}^{\dagger}U^{\dagger}A_{\alpha}$. This parametrization reduces
to the standard parametrization for $\alpha=0$ and has none of the flaws
of the naive parametrization.

The tree-level effective potential
$V_0={\cal H}_2^{\rm static}=-{\cal L}^{\rm static}$
is now evaluated to be
\begin{align}
V_0&=&-f^2B_0(2m\cos\alpha+m_s)-{1\over2}f^2\mu_I^2\sin^2\alpha\;.
\end{align}
At next-to-leading order in the low-energy expansion, there are
twelve operators. Not all of them are relevant for the present calculations,
in fact only eight contribute to the effective potential.
They are
\bqa\nonumber
{\cal L}_4&=& L_1\left({\rm Tr}
\left[\nabla_{\mu}\Sigma^{\dagger}\nabla^{\mu}\Sigma\right]\right)^2
\\ \nonumber&&
+L_2{\rm Tr}\left[\nabla_{\mu}\Sigma^{\dagger}\nabla_{\nu}\Sigma\right]
    {\rm Tr}\left[\nabla^{\mu}\Sigma^{\dagger}\nabla^{\nu}\Sigma\right]
\\ \nonumber&&
+L_3{\rm Tr}\left[(\nabla_{\mu}\Sigma^{\dagger}\nabla^{\mu}\Sigma)
(\nabla_{\nu}\Sigma^{\dagger}\nabla^{\nu}\Sigma)\right]
\\&& \nonumber
+L_4{\rm Tr}\left[\nabla_{\mu}\Sigma^{\dagger}\nabla^{\mu}\Sigma\right]
{\rm Tr}\left[\chi^{\dagger}\Sigma+\chi\Sigma^{\dagger}\right]
\\ \nonumber&&
+L_5{\rm Tr}\left[(\nabla_{\mu}\Sigma^{\dagger}\nabla^{\mu}\Sigma)
\left(\chi^{\dagger}\Sigma+\Sigma^{\dagger}\chi\right)\right]
\\ &&\nonumber
+L_6\left[{\rm Tr}\left(\chi^{\dagger}\Sigma+\chi
    \Sigma^{\dagger}\right)\right]^2
%\\ && \nonumber
%+L_7\left[{\rm Tr}\left(\chi^{\dagger}\Sigma-\chi\Sigma^{\dagger}\right)
%\right]^2
\\ && \nonumber
+L_8{\rm Tr}\left[\chi^{\dagger}\Sigma\chi^{\dagger}\Sigma
+  \chi\Sigma^{\dagger}\chi\Sigma^{\dagger}\right]
\\ &&+H_2{\rm Tr}[\chi^{\dagger}\chi]\;.
\label{lag}
\eqa
In writing the NLO Lagrangian above, we have ignored the Wess-Zumino-Witten
terms since they do not contribute to the quantities in the present paper.
The last term in Eq.~(\ref{lag})
is a contact term, which is needed to renormalize the
vacuum energy and to show the scale independence of the final result
for the effective potential in each phase.
The  contribution from the terms in Eq.~(\ref{lag})
to ${\cal H}_4^{\rm static}=-{\cal L}_4^{\rm static}=V_1^{\rm static}$ is
\bqa\nonumber
%{\cal H}_4^{\rm static}&=&
  V_{1}^{\rm static}
  &=& -(4L_1+4L_2+2L_3)\mu_I^4\sin^4\alpha
  \\ &&\nonumber
  -8L_4B_0(2m\cos\alpha+m_s)\mu_I^2\sin^2\alpha
  \\ &&\nonumber
  -8L_5B_0m\mu_I^2
  \cos\alpha\sin^2\alpha
  \\ &&\nonumber
  -16L_6B_0^2(2m\cos\alpha+m_s)^2  
  \\ &&\nonumber
  -  8L_8B_0^2(2m^2\cos2\alpha+m_s^2)
\\ &&
  -4H_2B_0^2(2m^2+m_s^2)\;.
\label{statpi}
\eqa
In a next-to-leading order calculation, we need to renormalize
the couplings $L_i$  and $H_i$ to eliminate the ultraviolet divergences that
arise from the functional determinants.
The relations between the  bare and renormalized couplings are
\bqa
\label{r1}
L_{i}&=&L_{i}^r(\Lambda)-
\Gamma_i\lambda\;,\\
H_i&=&H_i^r(\Lambda)-\Delta_i\lambda\;,
\label{r2}
\eqa
with $\lambda={\Lambda^{-2\epsilon}\over2(4\pi)^2}
\left[{1\over\epsilon}+1\right]$.
Here $\Gamma_i$ and $\Delta_i$ are constants~\cite{gasser2}
\begin{align}
&  \Gamma_{1}=\frac{3}{32}\;,& \Gamma_{2}&=\frac{3}{16}\;,& \Gamma_{3}&=0\;,
&  \Gamma_{4}={1\over8}\;,
  \\
    &  \Gamma_{5}=\frac{3}{8}\;,& \Gamma_{6}&=\frac{11}{144}\;,& 
\Gamma_{8}&={5\over48}\;,
&  \Delta_{2}={5\over24}\;,
%  &    \Delta_2  ={5\over24}    \;.
\end{align}
and $\Lambda$ is the
renormalization scale associated with the modified minimal
substraction scheme $\overline{\rm MS}$.
Taking the derivative of Eqs.~(\ref{r1})--(\ref{r2})
and using the fact that the bare couplings are scale independent,
one finds the renormalization group equations for the renormalized
couplings,
\bqa
\label{run1}
\Lambda{dL_i^r(\Lambda)\over d\Lambda}&=&-{\Gamma_i\over(4\pi)^2}\;,\\
\Lambda{dH_i^r(\Lambda)\over d\Lambda}&=&-{\Delta_i\over(4\pi)^2}\;.
\label{run2}
\eqa
The contact term $H_2{\rm Tr}[\chi^{\dagger}\chi]$ makes a constant contribution
to the effective potential which is independent of the chemical potential
and therefore the
same in both phases. We keep it, however, in the final expression
for the NLO effective potential since $H_2^r(\Lambda)$ is running. It is
needed to
show the scale independence of $V_{\rm eff}$.
The renormalized NLO effective potential
  $V_{\rm eff}=V_0+V_1+V_1^{\rm static}$ is given by %the sum of
%  Eqs.~(\ref{treepi}), ~(\ref{divpi}), and ~(\ref{statpi})
%  then reads
%  \bqa
  \bqa
    \nonumber
V_{\rm eff}&=&-f^2B_0(2m\cos\alpha+m_s)-{1\over2}f^2\mu_I^2\sin^2\alpha
\\ && \nonumber
-\bigg[4{L}_1^r+4{L}_2^r+2{L}_3^r
\bigg.\\ &&\left.\nonumber
  +{1\over16(4\pi)^2}\left({9\over2}+8\log{\Lambda^2\over m_3^2}
\right.\right.\\ && \nonumber\left.\left.
  +\log{\Lambda^2\over\tilde{m}_4^2}\right)
            \right]\mu_I^4\sin^4\alpha
\\ && \nonumber
-\left[8{L}_4^{r}
  +{1\over2(4\pi)^2}\left({1\over2}+\log{\Lambda^2\over\tilde{m}_4^2}\right)
\right]
\\&&\times \nonumber
                  B_0(2m\cos\alpha+m_s)\mu_I^2\sin^2\alpha
\\ && \nonumber
-\left[8{L}_5^{r}
  +{1\over2(4\pi)^2}\left({3\over2}+4\log{\Lambda^2\over m_3^2}
 -\log{\Lambda^2\over\tilde{m}_4^2}\right)     \right]
  \\&&\times \nonumber
  B_0m\mu_I^2\cos\alpha\sin^2\alpha
+B_0^2m^2\sin^2\alpha\left[16L_8^r-8H_2^r\right]
\\ && \nonumber
-\left[
  16L_6^r+8L_8^r+4H_2^r
  +{1\over(4\pi)^2}\left({13\over18}  +\log{\Lambda^2\over\tilde{m}_4^2}
\right.\right.\\ && \nonumber\left.\left.
+{4\over9}\log{\Lambda^2\over m_8^2}\right)\right]
               B_0^2m_s^2
\\ && \nonumber
-\left[  64L_6^r  +{1\over(4\pi)^2}\left({11\over9}
    +2\log{\Lambda^2\over\tilde{m}_4^2}
+{4\over9}\log{\Lambda^2\over m_8^2}  \right)\right]
\\&& \times \nonumber
     B_0^2mm_s\cos\alpha
\\ && \nonumber
-\left[
  64L_6^r+16L_8^r+8H_2^r    +{1\over(4\pi)^2}\left({37\over18}
+\log{\Lambda^2\over\tilde{m}_2^2}+
\right.\right. \\ &&\left.\left. \nonumber
     +2\log{\Lambda^2\over m_3^2}+
    \log{\Lambda^2\over\tilde{m}_4^2}
+{1\over9}\log{\Lambda^2\over m_8^2}
  \right)\right]B_0^2m^2\cos^2\alpha
\\ &&
+V_{\rm 1,{\pi^{+}}}^{\rm fin}
+V_{\rm 1,{\pi^{-}}}^{\rm fin}\;,
\label{renpi}
\eqa
%\eqa
where $L_i^r(\Lambda)$
are the renormalized coupling constants and the masses are
\bqa
\tilde{m}_2^2&=&2B_0m\cos\alpha\;,\\
m_3^2&=&2B_0m\cos\alpha+\mu_I^2\sin^2\alpha\;,\\
\tilde{m}_4^2&=&B_0(m\cos\alpha+m_s)+{1\over4}\mu_I^2\sin^2\alpha\;,\\
m_{8}^2&=&{2B_0(m\cos\alpha+2m_s)\over3}\;.
\eqa
%and $\Lambda$ is the renormalization scale associated
%with the modified minimal subtraction scheme $\overline{\rm MS}$.
Finally, 
$V_{1,{\pi^{\pm}}}^{\rm fin}$ %, $V_{\rm eff,{K^{\pm}}}^{\rm fin}$, and
%$V_{\rm eff,{K^{0}}}^{\rm fin}$
are finite subtraction terms  which depend on
  $B_{0}$ and $m$ but are independent of $m_{s}$. For details, see
  Ref.~\cite{moreresults}.
% The first line in Eq.~(\ref{renpi}) is from ${\cal L}_2^{\rm static}$,
%i.e. the order-$p^2$ term in $\chi$PT.
%The remaining terms are from ${\cal L}_4^{\rm static}$,
%as well as the bosonic functional determinants.
The couplings are running in such a way that their
$\Lambda$-dependence cancel against the explicit $\Lambda$-dependence
of the chiral logarithms
in Eq. (\ref{renpi}), implying that $\Lambda{dV_{\rm eff}\over d\Lambda}=0$,
cf. Eqs.~(\ref{run1})--(\ref{run2}).
%In the two-flavor case, the Lagrangian simplifies
%since there are fewer independent operators at order $p^4$.
%If we denote the bare couplings in the $SU(2)\times SU(2)$-case by $l_i$,
%one finds $l_1=4L_1+2L_3$, $L_2=4L_2$, $l_3=4(-2L_4-L_5+4L_6+2L_8)$, and
%$l_4=4(2L_3+L_5)$.
In order to obtain Eq.~(\ref{renpi}), we must
isolate the ultraviolet divergences from the functional determinants.
This is done by adding and subtracting a divergent term
that we calculate analytically in dimensional regularization. The subtracted
term is then combined with the original one-loop expression for the
effective potential giving finite terms $V_{1,\pi^{\pm}}^{\rm fin}$
that can be easily computed numerically.
The divergences are finally removed by renormalization of the $L_i$s
according to Eqs.~(\ref{r1})--(\ref{r2}).
The details of the subtraction and renormalization procedure can be
found in Ref.~\cite{moreresults}
and the NLO effective potential in the two-flavor case can be found
in Ref.~\cite{usagain}.

% The one-loop effective potential is
%\bqa\nonumber
%V_{\rm1}&=&{1\over2}\int_p\left[
%  E_{\pi^+}+E_{\pi^-}+E_{\pi^0}
%\right.\\ &&\left.
%  +  E_{K^+}+E_{K^-}+E_{K^0}+E_{\bar{K}^0}+E_{\eta^0}
%\right]\;,
%\eqa
%where the integral is defined using dimensional regularization
%in $d=3-2\epsilon$ dimensions,
%The large-$p$ behavior in Eq.~(\ref{expandp}) is the same
%as the sum $E_1+E_2$, where 
%$E_1=\sqrt{p^2+m_1^2+\mbox{$1\over4$}m_{12}^2}$ and
%$E_2=\sqrt{p^2+m_2^2+\mbox{$1\over4$}m_{12}^2}$.
%The integral over $E_{\pi^+}+E_{\pi^-}-E_1-E_2$ is therefore convergent
%in the ultraviolet
%and the subtraction integrals of $E_1$ and $E_2$ can be done analytically in
%dimensional regularization. %The result is given in Eq.~(\ref{int1}).
Thermodynamic quantities can be calculated from the effective potential
Eq.~(\ref{renpi}), for example the pressure $P=-V_{\rm eff}$, the
isospin density $n_I=-{\partial V_{\rm eff}\over\partial\mu_I}$, and the energy
density $\epsilon=-P+n_I\mu_I$. All these quantities are evaluated
at the value of $\alpha$ that minimizes the effective potential, i.e.
satifies ${\partial V_{\rm eff}\over\partial\alpha}=0$.
%At tree-level, it is easy to see that there are two solutions to
%this equation, namely $\alpha=0$ and $\alpha={\mu_I^2\over m^2}$.
%The former is the vacuum phase, while the latter is a phase of
%condensed charged pions. The critical isospin chemical potential
%$\mu_I^c=m$ which is identified with the pion mass at LO.
%At NLO, one can show that the transition is still located at $\mu_I^c=m_{\pi}$;
%however it is important to take the leading loop correction to the pion mass
%into account to show this explicitly. This result holds to all order
%in perturbation theory.

For sufficiently large values of $m_s$, we expect using effective-field
theory arguments,  that all degrees of freedom that contain an $s$-quark
freeze and decouple. Thus we expect that the kaons and eta decouple
from the low-energy dynamics involving the pions.
Formally, this is the limit $B_{0}m\ll B_{0}m_{s}\ll (4\pi f_{\pi})^{2}$.
The system is then
described in terms of two-flavor chiral perturbation theory where the
effects of the $s$-quark shows up in the renormalization of the
coupling constants $l_i$ of the form
$\log{\Lambda^2\over\tilde{m}_{K,0}^{2}}$
and $\log{\Lambda^2\over\tilde{m}_{\eta,0}^2}$,
where the masses are
$\tilde{m}_{K,0,}^2=B_0m_s$ and $\tilde{m}^2_{\eta,0}={4B_0m_s\over3}$.
Expanding the effective potential Eq.~(\ref{renpi}) in inverse powers
of $m_s$, we obtain
\bqa\nonumber
V_{\rm eff} &=& \nonumber
-2\tilde{f}^2\tilde{B}_0m\cos\alpha-f^2B_0m_s
-{1\over2}\tilde{f}^2\mu_I^2\sin^2\alpha
\\ &&\nonumber
-\left[4l_3^r+4l_4^r
  +{1\over(4\pi)^2}\left({3\over2}+\log{\Lambda^2\over\tilde{m}_1^2}
\right.\right.\\ &&\left.\left.\nonumber
    +2\log{\Lambda^2\over m_3^2}
\right)\right]
%\\ && \nonumber\times
  {B}_0^2m^2\cos^2\alpha
\nonumber\\ && \nonumber
-\left[l_4^r+{1\over(4\pi)^2}
  \left({1\over2}+\log{\Lambda^2\over m_3^2}
  \right)\right]
 \\ && \nonumber\times
  2{B}_0m\mu_I^2\cos\alpha\sin^2\alpha
\nonumber\\ &&
-\left[l_1^r+l_2^r+{1\over2(4\pi)^2}
  \left({1\over2}+\log{\Lambda^2\over m_3^2}
  \right)\right]
\mu_I^4\sin^4\alpha
 \nonumber\\ && \nonumber
\\ && \nonumber
+4(-h_1^r+l_4^r){B}_0^2m^2-\left[16L_6^r+8L_8^r+4H_2^r
\right.\\ && \nonumber\left.
  +{1\over(4\pi)^2}\left({13\over18}
  +\log{\Lambda^2\over\tilde{m}_{K,0}^2}
\right.\right.\\ && %\nonumber
\left.\left.
    +{4\over9}\log{\Lambda^2\over\tilde{m}_{\eta,0}^2}
\right)\right]
{B}_0^2%\\ &&\nonumber
%\times
%B_0^2
m_s^2
% \\ && 
 +V^{\rm fin}_{1,\pi^{+}}+V^{\rm fin}_{1,\pi^{-}}
 \;,
\label{yup}
 \eqa
where we have defined the combinations of the renormalized couplings
$l_i^r$ and $h_1^r$ as well as renormalized $\tilde{f}$ and $\tilde{B}_0$ as
\bqa\nonumber
l_1^r+l_2^r&=&4L_1^r+4L_2^r+2L_3^r
\\ && 
+{1\over16(4\pi)^2}
\left[\log{\Lambda^2\over\tilde{m}_{K,0}^2}-1\right]\;,
\label{rel1}
\\ \nonumber
l_3^r+l_4^r&=&
16L_6^r+8L_8^r+{1\over4(4\pi)^2}\left[
  \log{\Lambda^2\over\tilde{m}_{K,0}^2}-1\right]
\\ &&
+{1\over36(4\pi)^2}
\left[\log{\Lambda^2\over\tilde{m}_{\eta,0}^2}-1\right]
\;,\\ \nonumber
l_4^r&=&8L_4^r+4L_5^r
\nonumber
+{1\over4(4\pi)^2}\left[
  \log{\Lambda^2\over\tilde{m}_{K,0}^2}-1\right]\;,
\\
-h_1^r+l_4^r&=&4L_8^r-2H_2^r\;,
\label{rel4}\\ \nonumber
\tilde{f}^2&=&f^2\left[1+{B_0m_s\over f^2}
  \left(16L_4^r
\right.\right.\\ &&\left.\left.    
   +{1\over(4\pi)^2}\log{\Lambda^2\over\tilde{m}_{K,0}^2}\right)\right]\;,
\label{rel5}
\\  \nonumber
\tilde{B}_0&=&B_0\left[
  1-{B_0m_s\over f^2}\left(16L_4^r-32L_6^r
    -
\right.\right.\\ &&\left.\left.
    {2\over9(4\pi)^2}\log{\Lambda^2\over\tilde{m}_{\eta,0}^2}\right)
\right]\;.
\label{rel6}
\eqa
Several comments are in order: The terms in Eq.~(\ref{yup})
that are proportional to powers of $m_s$ are independent of $\alpha$
and $\mu_I$.
They can be interpreted as a constant renormalized contribution
to the vacuum energy from the $s$-quark and can be omitted.
The constant
term proportional to $B_0^2m^2$ can be omitted for similar reasons.
The relations between the renormalized couplings $l_i^r, h_i^r$ and the
low-energy constants $\bar{l}_i, \bar{h}_i$ in two-flavor $\chi$PT are
\bqa
l_i^r(\Lambda)&=&{\gamma_i\over2(4\pi)^2}
\left[\bar{l}_i+\log{2B_0m\over\Lambda^2}\right]\;,\\
h_i^r(\Lambda)&=&{\delta_i\over2(4\pi)^2}
\left[\bar{h}_i+\log{2B_0m\over\Lambda^2}\right]\;,
\label{lirlbare}
\eqa
where $\gamma_1={1\over3}$, $\gamma_2={2\over3}$, $\gamma_3=-{1\over2}$, 
$\gamma_4=2$, and $\delta_1=2$~\cite{gasser1}. The renormalization group
equations are
then $\Lambda{dl_i^r(\Lambda)\over d\Lambda}=-{\gamma_i\over(4\pi)^2}$.
Given the renormalization group equations for $l_i^r$, $h_i^r$, $L_i^r$,
$H_i^r$, one verifies that the $\Lambda$-dependence of the left - and
right-hand side in Eqs.~(\ref{rel1})--(\ref{rel4}) is identical.
Moreover, the parameters $\tilde{f}$ and $\tilde{B}$ are independent of
the scale.
Eqs.~(\ref{rel1})--(\ref{rel6}) are in agreement with the
original calculations of Ref.~\cite{gasser2}, where 
relations among the renormalized couplings in two - and three-flavor
$\chi$PT were derived.
This agreement is a nontrivial check of our calculations.
Inserting these relations using (\ref{lirlbare}) into Eq.~(\ref{yup}),
we finally obtain
\bqa\nonumber
V_{\rm eff}&=&-2\tilde{f}^{2}\tilde{B}_0m\cos\alpha-\frac{1}{2}\tilde{f}^{2}
\mu_{I}^{2}
\sin^{2}\alpha
\\ && \nonumber
-\frac{1}{(4\pi)^{2}}\left [\frac{3}{2}-\bar{l}_{3}
  +{4}\bar{l}_{4}
%  +%{1\over2}  \log\left({\Lambda^2\over\tilde{m}_1^2}\right)
  +\log\left({2B_0m\over \tilde{m}_1^2}\right)
\right.\\ &&\left. \nonumber
  +  2\log\left({2B_0m\over m_3^2}\right)
\right ]
%\right.\\ &&\nonumber\left.
%\\ && \nonumber\times
B_0^2m^2\cos^{2}\alpha
\\&& \nonumber
-\frac{1}{(4\pi)^{2}}
\left [{1\over2}+\bar{l}_{4}
%+\log\left({m_4^2\over\tilde{m}_1^2}\right)
  +  \log\left({2B_0m\over m_3^2}\right)
\right ]
%\right.\\ &&\left.
\\ &&\times \nonumber
2B_0m\mu_{I}^{2}\cos\alpha\sin^{2}\alpha
\\ && \nonumber
-\frac{1}{2(4\pi)^{2}}\left [{1\over2}
  +\frac{1}{3}\bar{l}_{1}+\frac{2}{3}\bar{l}_{2}
  %{1\over2}
%  \log\left({\lambda^2\over\tilde{m}_1^2}\right)
%\right.\\ &&\nonumber\left.
%  +\log\left({\Lambda^2\over \tilde{m}_2^2}\right)
  +  \log\left({2B_0m\over m_3^2}\right)
\right ]
\\ &&\times %\nonumber
\mu_{I}^{4}\sin^{4}\alpha
%\\ &&
%+V_{{\rm eff},\pi^0}^{\rm fin}
+V_{{\rm 1},\pi^+}^{\rm fin}+V_{{\rm 1},\pi^-}^{\rm fin}
\;.
\label{yup2}
\eqa
In the limit $B_{0}m_{s}\ll (4\pi f_{\pi})^{2}$, $B_{0}$ in the NLO terms can be
identified with $\tilde{B}_{0}$ using Eq.~(\ref{rel6}) and the result reduces to
that of two-flavor $\chi$PT in Ref.~\cite{usagain}.

\section{Results and discussion}
The expressions for the effective potential, isospin density, pressure, and
energy density are all expressed in terms of the isospin chemical potential,
the parameters $B_0m$, $B_0m_s$, and $f$ of the chiral Lagrangian
as well as the renormalized couplings $L_i^r$.
In order to make predictions, we need to determine the parameters
of the chiral Lagrangian using the physical meson masses and
  the decay constants.
In $\chi$PT, one can calculate the pole masses of the mesons and
the decay constants ($f_{\pi}$, $f_{K}$)
systematically in the low-energy expansion.
At one loop, the
results are expressed in
terms of $B_0m$, $B_0m_s$, $f$, and
$L_i^{r}$~\footnote{All the relevant
    relationships between bare and physical quantities (masses and decay
    constants) are stated in Ref.~\cite{moreresults}.}. These equations can be 
solved to find the parameters of the chiral Lagrangian and thereby
numerically evaluate the effective potential.
The tree-level values of $m_{\pi,0}$ and
$m_{K,0}$ can be expressed in terms of $B_0m$ and $B_0m_s$ as
$m_{\pi,0}^2=2B_0m$ and $m_{K,0}^2=B_0(m+m_s)$.
Since we want to compare our predictions with the results of the
lattice simulations~\cite{endro}, we
use their values for the meson masses and decay
  constants~\cite{private}, %for $m_{\pi}^2$, $m_K^2$, and$f_{\pi}$
\begin{align}
\label{masses}  
m_{\pi}&=131\pm3\;{\rm MeV}\;,
&m_{K}&=481\pm10\;{\rm MeV}\;,\\
%\nonumber
f_{\pi}&={128\pm3\over\sqrt{2}}{\rm MeV}\;,
&f_{K}&={150\pm 3\over\sqrt{2}}{\rm MeV}.
\label{decays}
\end{align}
The low-energy constants
have been determined
experimentally, with the following values and uncertainties
at the scale $\mu=m_{\rho}$, where $m_{\rho}$ is the mass of the $\rho$ meson and
$\Lambda^{2}=4\pi e^{-{\gamma}_E}\mu^{2}$~\cite{bijnensreview}
\begin{align}
%\nonumber
{L}_{1}^r&=(1.0\pm 0.1)\times10^{-3}
&{L}_{2}^r&=(1.6\pm 0.2)\times10^{-3}\\
%\nonumber
{L}_{3}^r&=(-3.8\pm 0.3)\times10^{-3}
&{L}_{4}^r&=(0.0 \pm 0.3)\times10^{-3}\\
{L}_{5}^r&=(1.2 \pm 0.1)\times10^{-3}
&{L}_{6}^r&=(0.0 \pm 0.4)\times10^{-3}\\
%\nonumber
%{L}_{7}^r&=(-0.4 \pm 0.2)\times10^{-3}
{L}_{8}^r&=(0.5 \pm 0.2)\times10^{-3}\;.&&
\label{LEC}
\end{align}
Since we need to determine three parameters in the effective
potential, we must choose three of the four
physical quantities from Eqs.~(\ref{masses})--(\ref{decays}).
For the results that we present below, we use
$m_{\pi}$, $m_K$, and $f_{\pi}$.
Using the one-loop $\chi$PT expression for
$f_K$, we obtain $f_K=113.9$ MeV for the central value, 
which is off by approximately 7\% compared to the lattice value
of $f_K={150\over\sqrt{2}}=106.1$ MeV.
The uncertainties in $L_i^r$, $m_{\pi}$, $m_K$, and $f_{\pi}$
translate into uncertainties in the parameters 
$B_0m$, $B_0m_s$, and $f$.
It turns out that the uncertainties in these parameters
in the three-flavor case are completely dominated by the uncertainties in
the LECs. In the two-flavor case, they are dominated by the
uncertainties in the pion mass and the pion decay constant. 
  Furthermore, for the lowest values of LECs obtained using the largest
  uncertainties in Eq.~(\ref{LEC}), the $\eta$ mass becomes imaginary and
  therefore unphysical. Consequently, we are forced to restrict the smallest
  value
 of the LECs used to ones obtained using $46\%$ of the total uncertainty.
We therefore simplify the analysis and add the uncertainties.
This yields
\begin{align}
m_{\pi,0}^{\rm cen}&=131.28\;{\rm MeV}\;
  &m_{K,0}^{\rm cen}=520.65\;{\rm MeV}\;\\
m_{\pi,0}^{\rm low}&=148.45\;{\rm MeV}\;
  &m_{K,0}^{\rm low}=617.35\;{\rm MeV}\;\\
  m_{\pi,0}^{\rm high}&=115.93\;{\rm MeV}\;
  &m_{K,0}^{\rm high}=437.84\;{\rm MeV}\;\\
  f^{\rm cen}&=75.16\;{\rm MeV}\;\\
  f^{\rm low}&=79.88\;{\rm MeV}\;\\
  f^{\rm high}&=70.44\;{\rm MeV}\;.  
\end{align}
Given that the effective potential derived in three-flavor
$\chi$PT of Eq.~(\ref{renpi}) reduces to the result in two-flavor $\chi$PT,
in the limit of light up and down quarks, it is worthwhile comparing
the predictions from two-flavor $\chi$PT from Ref.~\cite{usagain}
using the $N_f=2$ LECs from the literature 
and those obtained by using Eqs.~(\ref{rel1})--(\ref{rel4}).
The $N_{f}=2$ LECs have the following values~\footnote{We note
    that it is standard practice to quote the LECs in two-flavor $\chi$PT using
    $\bar{l}_{i}$ defined through Eq.~(\ref{lirlbare}). On the other hand, for
    three-flavor $\chi$PT, quoting $L_{i}^{r}$ at the scale $\mu$ equal to the
    $\rho$ mass ($m_{\rho}$) is standard.}~\cite{bijnensreview}
\begin{align}
\label{LEC2old}
\bar{l}_{1}(N_{f}=2)&=-0.4&\bar{l}_{2}(N_{f}=2)&=4.3\\
\bar{l}_{3}(N_{f}=2)&=2.9&\bar{l}_{4}(N_{f}=2)&=4.4\;.
\end{align}
The three-flavor LECs $L_i^r$ are the running couplings
  evaluated at the scale $m_{\rho}$ and we use their renormalization group
  equations to run them to the scale $m_{\pi,0}$, where
  the two-flavor LECs ($\bar{l}_i$), defined in Eq.~(\ref{lirlbare}),
are evaluated according to Eqs.~(\ref{rel1})--(\ref{rel4}).
We then get the following central
values
\begin{align}
\label{newLEC}
\bar{l}_{1}(N_{f}=3)&=14.5&\bar{l}_{2}(N_{f}=3)&=6.5\\
  \bar{l}_{3}
  (N_{f}=3)&=4.1&\bar{l}_{4}(N_{f}=3)&=4.2\;.
\end{align}
%\begin{equation}
%\begin{split}
%\label{newLEC}
%l^{r}_{1}(N_{f}=3)&=0.01157\\
%l^{r}_{2}(N_{f}=3)&=0.006343\\
%l^{r}_{3}(N_{f}=3)&=-0.0008860\\
%l^{r}_{4}(N_{f}=3)&=0.004456\ .
%\end{split}
%\end{equation}
\begin{figure}[htb]
  \includegraphics[width=0.45\textwidth]{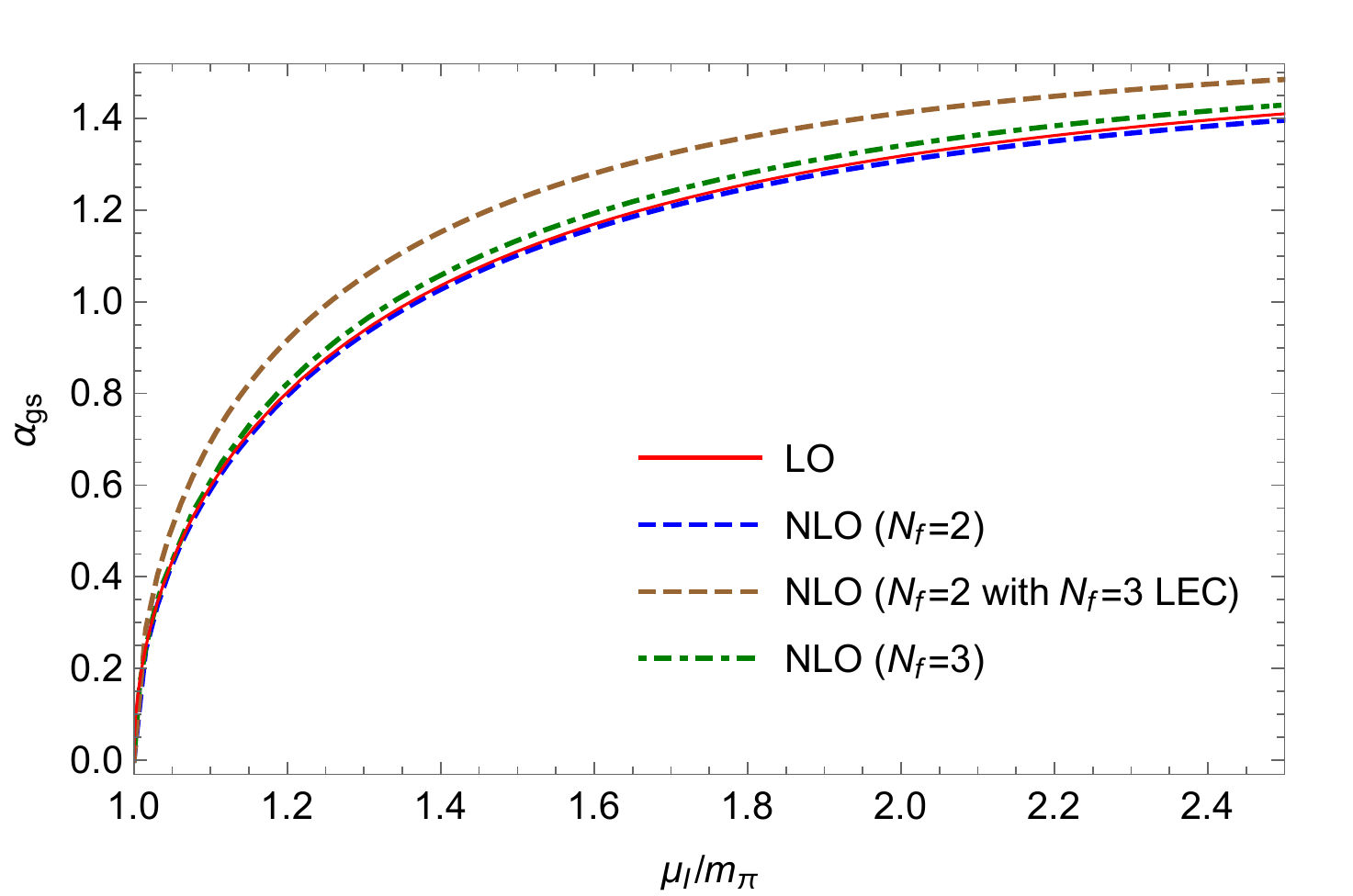}
  \caption{$\alpha_{\rm gs}$ as a function of $\mu_I/m_{\pi}$ at LO (red), at
    NLO with two flavors (blue), NLO with three flavors (green), and
   NLO with two flavors and three-flavor LECs (brown).
  See main text for details.}
\label{al}
\end{figure}
The disagreement is most significant in $\bar{l}_{1}$ which
have signs that are opposite in the two-flavor versus
the three-flavor case. The differences in the other LECs are less significant
but still
non-trivial except for $\bar{l}_{4}$. In order to evaluate the effect of these
discrepancies on physical observables in the pion-condensed phase, we have
generated the isospin density, pressure, and the equation of state using the
two-flavor LEC values generated using three-flavor LECs,
which we discuss at the end of this section.

The equation ${\partial V_{\rm eff}\over\partial\alpha}=0$ has two types of
solutions.
For $\mu_I<m_{\pi}$, the solution is $\alpha=0$, where it is
straightforward
to show that the effective potential and therefore the thermodynamic
functions are independent of $\mu_I$. We refer to this
phase as the vacuum phase, which exhibits the Silver Blaze
property~\cite{cohen}, namely that the thermodynamic functions
are independent of $\mu_I$ up to a critical value $\mu_{I}^{c}=m_{\pi}$.
For $\mu_I>m_{\pi}$, we have a nonzero condensate of $\pi^+$, which
breaks the $U(1)_{I_3}$ symmetry of the chiral Lagrangian, and a nonzero
value for $\alpha$. In Fig.~\ref{al}, we show the solution $\alpha_{\rm gs}$
to the equation
${\partial V_{\rm eff}\over\partial\alpha}=0$ as a function of
$\mu_I\over m_{\pi}$
  at LO.~\footnote{At LO, the two and three-flavor results for
    $\alpha$  coincide.}
For asymptotically large values of the isospin chemical,
$\alpha_{\rm gs}$ approaches ${\pi\over2}$.

We next expand the effective potential around $\alpha=0$
to obtain a Ginzburg-Landau energy functional that can be used to
determine the order of the phase transition.  
This expansion is valid close to the phase transition where $\alpha\ll1$.
To fourth-order, we obtain 
\bqa
V_{\rm eff}^{\rm LG}=a_0(\mu_I)+a_2(\mu_I)\alpha^2+a_4(\mu_I)\alpha^4\;.
\eqa
The vanishing of $a_2$ defines the critical chemical potential $\mu_I^c$. Since
$a_2=f_{\pi}^2(\mu_I^2-m_{\pi}^2)$, we have $\mu_I^c=m_{\pi}$.
The onset of Bose condensation at $\mu_I^c=m_{\pi}$ is an exact result.
Moreover, since the coefficient
$a_4(\mu_I^c)>0$, the transition to a pion-condensed phase is of second order,
with mean field critical exponents.
These results are in agreement with lattice
simulations~\cite{gergy1, gergy2, gergy3} as well as model
calculations~\cite{allofus}.

\begin{figure}[htb]
  \includegraphics[width=0.45\textwidth]{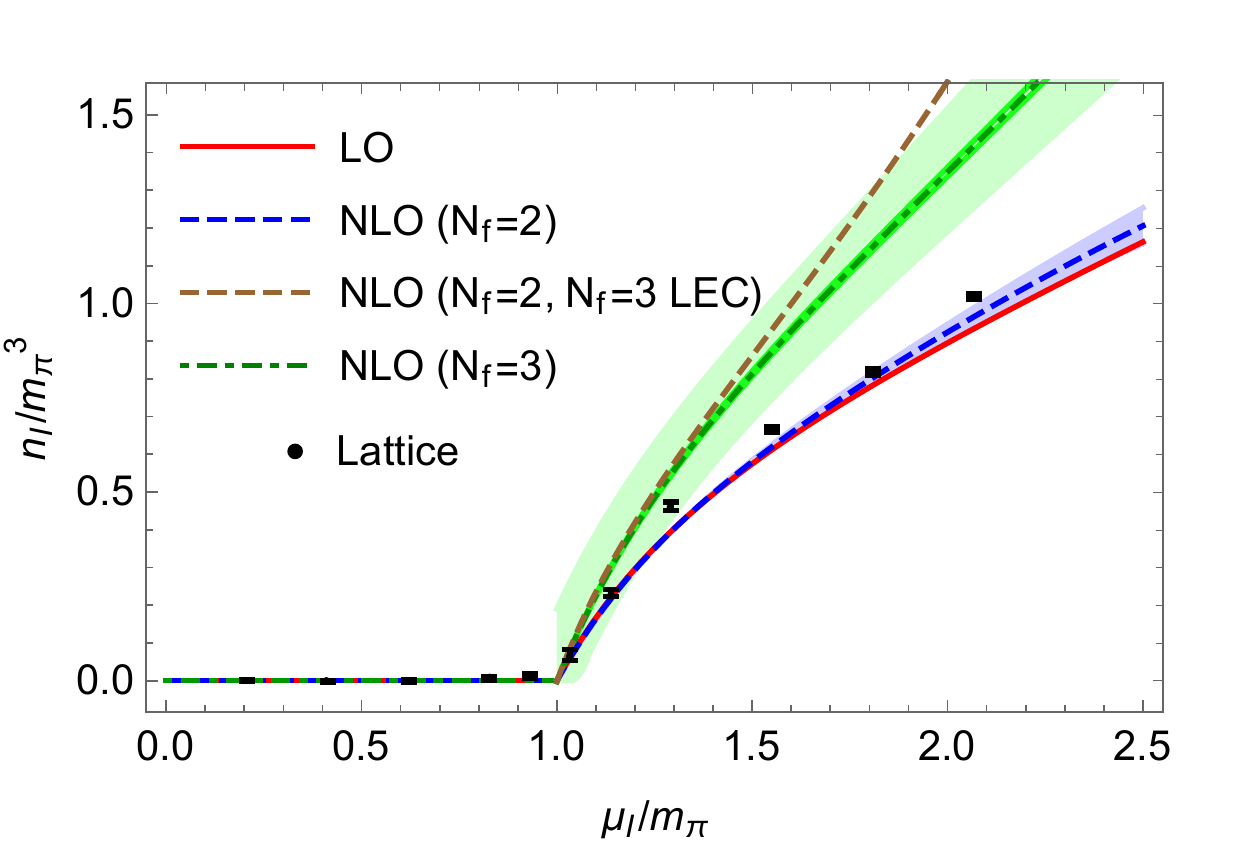}
\caption{Normalized isospin density as a function of $\mu_I/m_{\pi}$
    at LO (red), at
    NLO with two flavors (blue), NLO with three flavors (green), and
       NLO with two flavors and three-flavor LECs (brown).
    See  main text for details.}
\label{iso}
\end{figure}

In Fig.~\ref{iso}, we show the isospin $n_I$ divided by $m_{\pi}^3$
as a function of $\mu_I/m_{\pi}$. The red solid line is the LO result.
Note that the LO result is the same in the two and three-flavor cases
for all thermodynamic quantities.
We have used the central values for
the low-energy constants $\bar{l}_i$ in the two-flavor case
to obtain the blue dashed line as explained in Ref.~\cite{usagain}. 
The blue band is obtained by including their uncertainties.
The light green band is the result 
of the three-flavor calculation with the minimum, central, and maximum
values of the parameters discussed above, while the dark
green band is from using the
central values of $L_i^{r}$ with uncertainties coming from the
lattice parameters only.

  The data points shown in Fig.~\ref{iso} are from the lattice calculations of
Refs.~\cite{gergy1,gergy2,gergy3}.
The two-flavor band is very small compared to the three-flavor band reflecting
the large uncertainty in the three-flavor $L_i^{r}$s. The
central line in the
three-flavor case is in very good agreement with lattice data
up to approximately $\mu_I\sim200$ MeV. After this, the curve overshoots
and for larger values the two-flavor central curve is in
much better agreement with lattice data.

\begin{figure}[htb]
\includegraphics[width=0.45\textwidth]{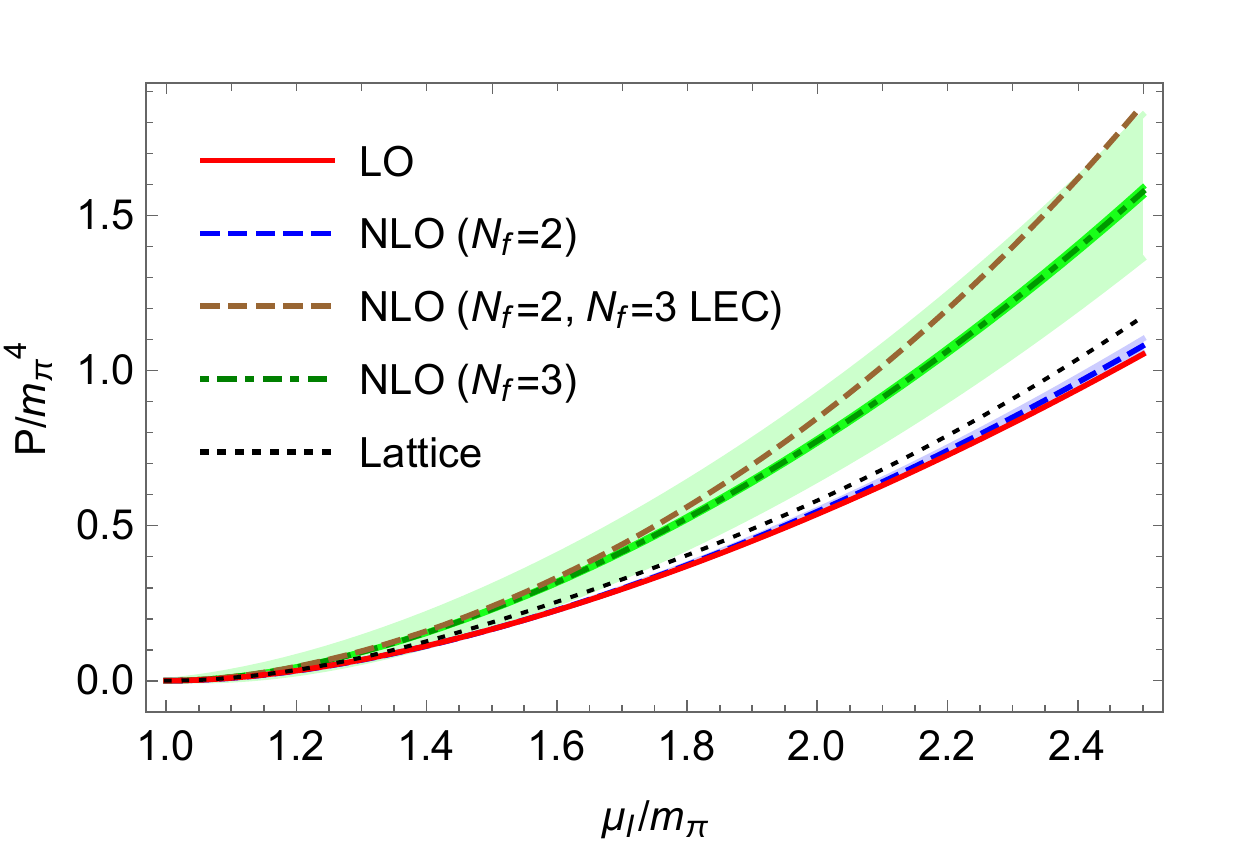}
  \caption{Pressure normalized by $m_{\pi}^4$ as a function of $\mu_I/m_{\pi}$
    at LO (red), at
    NLO with two flavors (blue), NLO with three flavors (green), and 
           NLO with two flavors and three-flavor LECs (brown).
    See  main text for details.}
  \label{presieure}
  \end{figure}

In Fig.~\ref{presieure}, we show the pressure $P$ divided by $m_{\pi}^4$
as a function of $\mu_I/m_{\pi}$. Note that we have subtracted the
pressure in the vacuum phase which is given by evaluating the
  negative of Eq.~(\ref{renpi})
for $\alpha=0$.
The red line is the LO result.
The blue dashed line is again
the result from two-flavor $\chi$PT using the central
values of $\bar{l}_i$, while the
band is obtained by including their uncertainties.
Similarly, the dashed-dotted line corresponds to the
central values of the $L^{r}_i$s in the
three-flavor case, while the light green band is
obtained by including their uncertainties.
Finally, by including only the uncertainties from the lattice parameters
we obtain the much narrower dark green band.
Here, the LO and the two-flavor results very close
in the entire range and systematically slightly
below the lattice data.
The three-flavor curve is in very good agreement with the results
of the Monte Carlo simulations up to $\mu_I=200$ MeV, after which
it overestimates the pressure.
\begin{figure}[htb]
  \includegraphics[width=0.45\textwidth]{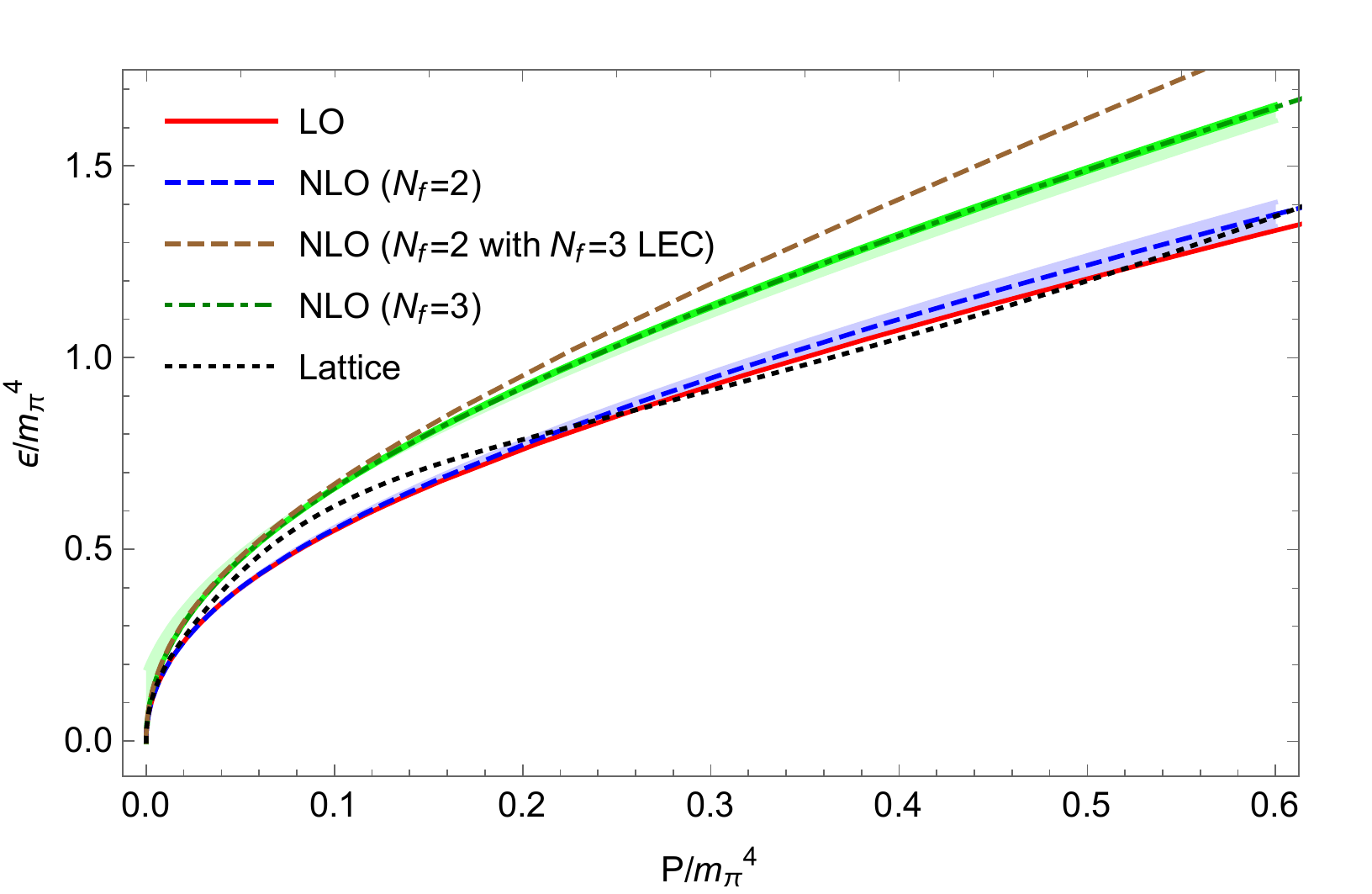}
  \caption{Energy density as a function of pressure, both normalized by
    $m_{\pi}^4$, at LO (red), at
    NLO with two flavors (blue), NLO with three flavors (green), and 
           NLO with two flavors and three-flavor LECs (brown).
    See main text for details.}
  \label{eoss}
\end{figure}

In Fig.~\ref{eoss}, we show the energy density $\epsilon$ divided by $m_{\pi}^4$
as a function of pressure $P$ divided by $m_{\pi}^4$. For all
  values of $P\over m_{\pi}^4$ three-flavor $\chi$PT overestimates the energy
  density compared to lattice data though for values of $P\over m_{\pi}^4$ up to
  approximately $0.10$, the discrepancy is quite small. On the other hand,
  two-flavor $\chi$PT underestimates the energy density as a function of
  pressure for values of $P\over m_{\pi}^4$ up to $0.20$. For values larger than
approximately $0.20$, two-flavor $\chi$PT agrees very well with lattice results.

Given the results shown in Figs. \ref{iso}, \ref{presieure} and \ref{eoss}
above, in particular the large differences between the results in two-flavor
and three-flavor $\chi$PT and the results in lattice QCD compared to
three-flavor $\chi$PT, it is important to explain this large discrepancy. The
naive expectation is that the loop effects from the strange quarks in
three-flavor $\chi$PT are small since the effect is sub-leading in the chiral
expansion. Furthermore, their effects should be suppressed since strange quark
masses are considerably larger than the masses of the up and down quarks. While
this picture is correct, it ignores the significant differences between the low
energy constants of two-flavor $\chi$PT and the ones that are extracted from
three-flavor $\chi$PT after integrating out the effect of the strange quarks.
We list the values in Eqs.~(\ref{LEC2old}) and (\ref{newLEC}) noting
significant discrepancies between the two sets. In each of the
figures (\ref{iso}, \ref{presieure} and \ref{eoss}), we incorporate an
additional result in two-flavor $\chi$PT using three-flavor LECs shown using
brown and dashed lines. We note that even two-flavor $\chi$PT using
three-flavor LECs overestimates the isospin density, pressure and the energy
density compared to lattice QCD results. For isospin chemical potential near
the second order phase transition up to approximately
$\frac{\mu_{I}}{m_{\pi}}\sim1.3$, the differences in the LECs fully explains the
discrepancy. For larger values of isospin chemical potential, the role of
strange quark loops becomes more significant -- our results suggests that they
have a negative effect on the pressure and isospin density compared to the
effects of the up and down quarks.

\section{Acknowledgments}
The authors would like to thank B. Brandt, G. Endr\H{o}di, and S. Schmalzbauer
for useful discussions as well as for providing the
data points of Ref.~\cite{endro}.

\bibliography{refs}
\end{document}